\documentclass[reqno,11pt]{amsart}

\usepackage[utf8]{inputenc}
\usepackage[english]{babel}
\usepackage[T1]{fontenc}
\usepackage[left=2.5cm, right=2.5cm, top=2.5cm, bottom=2.5cm]{geometry}
\usepackage{datetime}
\usepackage{lineno}
\usepackage{amsmath,amssymb,amsthm}
\usepackage[dvipsnames]{xcolor}
\usepackage{tikz}
\usepackage{graphicx}
\usepackage{subcaption}
\usepackage{enumerate} 
\usepackage{url,hyperref}
\usepackage[noend]{algpseudocode}
\usepackage{algorithm}
\usepackage{doi}

\usepackage{tikz}
\usepackage{caption, subcaption}

\newcommand{\up}[1]{\raisebox{1.5ex}[0pt]{#1}}

\newcommand{\defi}[1]{\emph{\color{red!50!black}#1}}

\newcommand{\CommentHigh}[1]{\Comment{\textrm{\textit{#1}}}}
\newcommand{\var}[1]{\textsf{#1}}

\def\orcidID#1{\unskip$^{[#1]}$}

\newcommand*\patchAmsMathEnvironmentForLineno[1]{%
\expandafter\let\csname old#1\expandafter\endcsname\csname #1\endcsname
\expandafter\let\csname oldend#1\expandafter\endcsname\csname end#1\endcsname
\renewenvironment{#1}%
{\linenomath\csname old#1\endcsname}%
{\csname oldend#1\endcsname\endlinenomath}}%
\newcommand*\patchBothAmsMathEnvironmentsForLineno[1]{%
\patchAmsMathEnvironmentForLineno{#1}%
\patchAmsMathEnvironmentForLineno{#1*}}%
\AtBeginDocument{%
\patchBothAmsMathEnvironmentsForLineno{equation}%
\patchBothAmsMathEnvironmentsForLineno{align}%
\patchBothAmsMathEnvironmentsForLineno{flalign}%
\patchBothAmsMathEnvironmentsForLineno{alignat}%
\patchBothAmsMathEnvironmentsForLineno{gather}%
\patchBothAmsMathEnvironmentsForLineno{multline}%
}

\newtheorem{theorem}{Theorem}
\newtheorem{lemma}[theorem]{Lemma}

\newtheorem*{question*}{Question}

\hypersetup{%
    colorlinks,%
    linkcolor = {red!60!black},%
    citecolor = {green!60!black},%
    urlcolor = {blue!60!black}%
}

\title[A faster heuristic for the Traveling Salesman Problem with Drone]{A faster heuristic for the \\Traveling Salesman Problem with Drone}

\author[Hokama]{Pedro H. D. B. Hokama\orcidID{0000-0002-3734-7679}}
\address[P. H. D. B. Hokama]{
  Instituto de Matemática e Computação -- 
  Universidade Federal de Itajubá -- 
  Brazil
}

\author[Lintzmayer]{Carla N. Lintzmayer\orcidID{0000-0003-0602-6298}}
\address[C. N. Lintzmayer]{
  Centro de Matemática, Computação e Cognição -- 
  Universidade Federal do ABC -- 
  Brazil
}

\author[San Felice]{Mário C. San Felice\orcidID{0000-0001-6476-4534}}
\address[M. C. San Felice]{
  Departamento de Computação --
  Universidade Federal de São Carlos -- 
  Brazil
}

\email{hokama@unifei.edu.br, carla.negri@ufabc.edu.br, felice@ufscar.br}

\thanks{
  This research has been partially supported by Coordenação de Aperfeiçoamento
  de Pessoal de Nível Superior -- Brasil -- CAPES -- Finance Code 001, and by 
  CNPq grant 404315/2023-2.
  C. N. Lintzmayer is partially supported by CNPq (Proc.~312026/2021-8) and by
  L'ORÉAL-UNESCO-ABC For Women In Science.
  CNPq is the National Council for Scientific and Technological Development of
  Brazil.
}

\begin{document}

\maketitle

\begin{abstract}
    Given a set of customers, the \defi{Flying Sidekick Traveling Salesman Problem (FSTSP)} consists of using one truck and one drone to perform deliveries to them.
    The drone is limited to delivering to one customer at a time, after which it returns to the truck, from where it can be launched again.
    The goal is to minimize the time required to service all customers and return both vehicles to the depot.
    In the literature, we can find heuristics for this problem that follow the order-first split-second approach: find a Hamiltonian cycle~$h$ with all customers, and then remove some customers to be handled by the drone while deciding from where the drone will be launched and where it will be retrieved.
    Indeed, they optimally solve the \defi{h-FSTSP}, which is a variation that consists of solving the FSTSP while respecting a given initial cycle~$h$.
    
    We present the Lazy Drone Property, which guarantees that only some combinations of nodes for launch and retrieval of the drone need to be considered by algorithms for the h-FSTSP.
    We also present an algorithm that uses the property, and we show experimental results which corroborate its effectiveness in decreasing the running time of such algorithms.
    Our algorithm was shown to be more than 84 times faster than the previously best-known ones over the literature benchmark.
    Moreover, on average, it considered a number of launch and retrieval pairs that is linear on the number of customers, indicating that the algorithm's performance should be sustainable for larger instances.
\end{abstract}

%\begin{CCSXML}
%<ccs2012>
%   <concept>
%       <concept_id>10010405.10010481.10010485</concept_id>
%       <concept_desc>Applied computing~Transportation</concept_desc>
%       <concept_significance>300</concept_significance>
%       </concept>
%   <concept>
%       <concept_id>10003752.10003809.10003635</concept_id>
%       <concept_desc>Theory of computation~Graph algorithms analysis</concept_desc>
%       <concept_significance>500</concept_significance>
%       </concept>
% </ccs2012>
%\end{CCSXML}
%
%\ccsdesc[300]{Applied computing~Transportation}
%\ccsdesc[500]{Theory of computation~Graph algorithms analysis}

%%%%%%%%%%%%%%%%%%%%%%%%%%%%%%%%%%%%%%%%%%%%%%%%%%%%%%%%%%%%%%%%%%%%%%%%%%%%%%%
\section{Introduction}

The problem we consider in this paper belongs to a large class of delivery problems which
involve using drones along with trucks to perform deliveries.
We specifically consider the case where there is one truck and one drone available, and either
of them can perform deliveries.
Within this case, the two main known variants are the \defi{Flying Sidekick Traveling Salesman 
Problem} (\defi{FSTSP}), defined by Murray and Chu~\cite{MurrayC2015}, and 
the \defi{Traveling Salesman Problem with Drone} (\defi{TSP-D}), defined 
by Agatz \textit{et al.}~\cite{AgatzBS2018}.
We deal with the former, and the main difference between the two is that the latter
allows for the truck and drone to visit a same node more than once.

The general idea of the FSTSP is that the truck will perform some route to deliver goods
to a subset of the customers (leaving from and returning to a depot), while the drone will be
launched from the truck to deliver goods to the remaining customers.
The drone, however, is limited to delivering to one customer at a time.
Once the drone delivers the parcel to one customer, it must return to the truck, from where
it can be launched again.
The goal is to minimize the time required to service all customers and return both vehicles
to the depot.

As generalizations of the classic Traveling Salesman Problem (TSP), these problems are
also NP-hard.
Several algorithms have been proposed to handle both FSTSP and TSP-D, and we are interested
in those that fall into the category of \defi{order-first split-second} heuristics:
the first step is building a Hamiltonian cycle on all customers (to be handled by the truck only), and 
the second step is removing some customers from this cycle to be handled by the drone.
Particularly, the order of the nodes visited by the truck respects the relative order of
the nodes in the initial cycle.

Murray and Chu~\cite{MurrayC2015} gave a mixed integer linear programming formulation 
and a route and re-assign heuristic for the FSTSP.
Their heuristic starts from a Hamiltonian cycle but not necessarily keeps the order of visits according 
to it.
Agatz \textit{et al.}~\cite{AgatzBS2018} presented an integer programming formulation and 
an order-first split-second heuristic for the TSP-D.
Interestingly, they compared their heuristic with the one from Murray and Chu and showed their
has some advantages indeed.

The splitting part of Agatz \textit{et al.}'s heuristic is a dynamic programming algorithm that provides an optimal solution for what we call the \defi{h-FSTSP}: a problem that consists of solving the FSTSP while respecting a given initial Hamiltonian cycle~$h$. See Figure~\ref{fig:instance} for an example.
In fact, the splitting part of all known order-first split-second heuristics for the FSTSP actually solve the h-FSTSP.

\begin{figure}[h]
\begin{center}
\begin{subfigure}[b]{0.40\textwidth}
\centering
\resizebox {1\textwidth} {!} {
\begin{tikzpicture}[scale=0.5,dot/.style={draw,fill,circle,inner sep=30pt},dep/.style={draw,fill,inner sep=40pt}]
  \tiny
  \draw (-20,-50) rectangle (230,100);

    \node[dep] (v0) at (-9.237203856626833, -12.024194375982791) {}; 
    \node[dot] (v1) at (215.41696051935992, 10.294392574871816){}; 
    \node[dot] (v2) at (1.5272180996244364, -5.295791741713567){}; 
    \node[dot] (v3) at (156.81385686558473, 84.48262742416554){}; 
    \node[dot] (v4) at (199.38876101383798, 16.39479379832618){}; 
    \node[dot] (v5) at (188.74022232542464, -4.624791540373848){}; 
    \node[dot] (v6) at (15.962689831849707, 24.390675185083044){}; 
    \node[dot] (v7) at (211.5246053740201, -27.14548195475132){}; 
    \node[dot] (v8) at (63.4842355059337, -24.88512533968853){}; 

    \draw (v0) -- (v8);
    \draw (v8) -- (v7);
    \draw (v7) -- (v5);
    \draw (v5) -- (v4);
    \draw (v4) -- (v1);
    \draw (v1) -- (v3);
    \draw (v3) -- (v6);
    \draw (v6) -- (v2);
    \draw (v2) -- (v0);

%\def\curv{10} %esse valor, quanto maior mais redondinho, 
%  \path [bend left=\curv] (Np) edge (Nd);
%  \path [bend left=25] (Op) edge (Od);
%  \path [bend left=\curv] (Pp) edge (Pd);

\end{tikzpicture}
}
\caption{Initial Hamiltonian cycle\label{fig:instance_lkh}}
\end{subfigure}
\begin{subfigure}[b]{0.40\textwidth}
\centering
\resizebox {1\textwidth} {!} {
\begin{tikzpicture}[scale=0.5,dot/.style={draw,fill,circle,inner sep=30pt},dep/.style={draw,fill,inner sep=40pt}]
  \tiny
  \draw (-20,-50) rectangle (230,100);

    \node[dep] (v0) at (-9.237203856626833, -12.024194375982791) {}; 
    \node[dot] (v1) at (215.41696051935992, 10.294392574871816){}; 
    \node[dot] (v2) at (1.5272180996244364, -5.295791741713567){}; 
    \node[dot] (v3) at (156.81385686558473, 84.48262742416554){}; 
    \node[dot] (v4) at (199.38876101383798, 16.39479379832618){}; 
    \node[dot] (v5) at (188.74022232542464, -4.624791540373848){}; 
    \node[dot] (v6) at (15.962689831849707, 24.390675185083044){}; 
    \node[dot] (v7) at (211.5246053740201, -27.14548195475132){}; 
    \node[dot] (v8) at (63.4842355059337, -24.88512533968853){};

    \draw (v0) -- (v8);
    \draw (v8) -- (v5);
    \draw (v5) -- (v1);
    \draw (v1) -- (v6);
    \draw (v6) -- (v0);

    \begin{scope}   [dashed,dash pattern=on 10pt off 20pt] 
        \draw (v0) -- (v7);
        \draw (v7) -- (v5);

        \draw (v5) -- (v4);
        \draw (v4) -- (v1);

        \draw (v1) -- (v3);
        \draw (v3) -- (v6);
        
        \draw (v6) -- (v2);
        \draw (v2) -- (v0);
    \end{scope}
    
\end{tikzpicture}
}
\caption{Optimum solution for h-FSTSP \label{fig:hfstsp}}
\end{subfigure}
\end{center}
\vspace{-0.5cm}
\caption{Instance with 8 clients, where the dotted lines represent the drone path.
}
\label{fig:instance}
\end{figure}

Other works considered the FSTSP and TSP-D, as well as some variants, presenting metaheuristics 
or improvements on exact formulations and exact approaches. 
We refer the reader to the nice survey by Macrina \textit{et al.}~\cite{MacrinaDGL2020} for more
details on them.
For our paper, though, it is worth highlighting two of these works.
Ha \textit{et al.}~\cite{HaDPH2018} proposed an order-first split-second heuristic for a variation
of the FSTSP in which the objective function is operational costs instead of completion time.
The splitting part of their heuristic is a graph algorithm called \defi{Split Algorithm}, and they showed that it solves the h-FSTSP.
Kundu \textit{et al.}~\cite{KunduEM2022} also dealt with the FSTSP (original objective function)
by showing an order-first split-second heuristic very similar to the Split Algorithm of 
Ha \textit{et al.}, but with some modifications that improve the empirical results, being the
one with the currently fastest results in the literature for the h-FSTSP.
It is important to notice that these algorithms are conceptually similar to the one of Agatz \textit{et al.}

In this paper we prove the Lazy Drone Property, which guarantees that only some combinations of nodes for launch and retrieval of the drone need to be considered by algorithms that solve the h-FSTSP.
We also present algorithms for the h-FSTSP, inspired by the Split Algorithm, that uses the Lazy Drone Property to make fewer comparisons when considering which nodes should be served by drones, while still solving optimally the h-FSTSP.
Moreover, we present empirical results using the literature benchmark to corroborate the greater efficiency of our algorithms when compared to the best-known in the literature.

The rest of the paper is structured as follows.
Section~\ref{sec:preliminiaries} presents some important notation and the formal definition
of the FSTSP and the h-FSTSP.
Section~\ref{sec:split} presents the Split Algorithm, as proposed by Ha \textit{et al.}~\cite{HaDPH2018}.
Section~\ref{sec:lazy_property} proves the Lazy Drone Property, while Section~\ref{sec:split_lazy} shows our Split Lazy Algorithm.
Section~\ref{sec:computational} presents the results of distinct implementations of our algorithm when running over
the benchmark from Bouman \textit{et al.}~\cite{BoumanAS2018}, as well as compares it 
with three existing algorithms.
At last, Section~\ref{sec:remarks} gives some final remarks.

%%%%%%%%%%%%%%%%%%%%%%%%%%%%%%%%%%%%%%%%%%%%%%%%%%%%%%%%%%%%%%%%%%%%%%%%%%%%%%%
\section{Preliminaries and notation}
\label{sec:preliminiaries}

In this section, we formally define the FSTSP.
We follow the definition given by Murray and Chu~\cite{MurrayC2015} and consider some
simplifications that arise from the benchmark we use.
For some integer $n > 0$, let $C = \{1, 2, \ldots, n\}$ be the set of all \defi{customers}, 
$N = \{0,1,\ldots,n+1\}$ be the set of all \defi{nodes} in the network where $0 \equiv n+1$ 
is the \defi{depot},
$N_D = \{0,1,\ldots,n\}$ be the set of nodes from which a vehicle may depart, 
$N_V = \{1,2,\ldots,n+1\}$ be the set of nodes that a vehicle may visit during a tour,
$t_R(i, j)$ be the \defi{truck time} to travel from $i \in N_D$ to $j \in N_V$,
and $t_D(i, j)$ be the analogous \defi{drone time}.
We have the following restrictions and assumptions:
\begin{itemize}
    \item each customer must be served exactly once by either the truck or the drone;
    \item the truck and the drone must depart from and return to the depot exactly once;
    \item the drone may make multiple \defi{sorties}, each consisting of three locations 
    $(i,j,k)$ where $i \in N_D$, $j \in C$, and $k \in N_V$, such that $i \neq j$, 
    $j \neq k$, and $k \neq i$. 
    We call~$i$ the \defi{launch node} and~$k$ the \defi{rendezvous node};
    \item the truck may visit multiple customers while the drone is in flight;
    \item the drone can be collected by the truck at some node~$i$ and be re-launched from~$i$,
    but if the drone is launched at~$i$, it may not return to the truck at node~$i$.
\end{itemize}
A node that is visited by the truck will be called a \defi{truck node}, while a node that is
only visited by the drone will be called a \defi{drone node}.

We consider a solution to the problem divided into smaller parts called operations.
An \defi{operation}~$o$ consists of a pair $(r,d)$ such that~$r$ and~$d$ are sequences of nodes,
where the former is associated with the truck while the latter is associated with the drone.
The sequence~$r = (v^r_1, \ldots, v^r_{|r|})$ describes the \defi{truck path} from the 
start node~$v^r_1 \in N_D$ to the final node~$v^r_{|r|} \in N_V$, which means all~$v^r_i$ 
are truck nodes and $|r| \geq 2$.
The time (cost) of such sequence, denoted by $t(r)$, is defined as 
\begin{equation*}
    t(r) = \sum_{\ell=1}^{|r|-1} t_R(v_{\ell}, v_{\ell+1}) \enspace .
\end{equation*}
The sequence~$d$ might be empty, indicating that the truck is carrying the drone during 
operation~$o$.
If~$d$ is not empty, then it describes the \defi{drone path}, which is a sequence of three nodes 
with the first and last coinciding with those from~$r$, that is, $d = (v^r_1, v^d, v^r_{|r|})$
such that $v^d \in C$ is a drone node.
The time (cost) of such sequence, denoted by $t(d)$, is defined as 
\begin{equation*}
    t(d) = t_D(v^r_1, v^d) + t_D(v^d, v^r_{|r|}) \enspace .
\end{equation*}
At last, the time of operation $o = (r,d)$, denoted by $t(o)$, is defined as
\begin{equation*}
    t(o) = \begin{cases}
        \max\left\{t(r), \, t(d)\right\} & \text{ if } d \text{ is not empty,} \\
        t(r) & \text{ otherwise.}
    \end{cases}
\end{equation*}

A \defi{solution} for the FSTSP consists of a sequence $S = (o_1, \ldots, o_{|S|})$ of operations
where each $o_\ell = (r_\ell, d_\ell)$, the first node of~$r_1$ is the depot, the last node 
of~$r_{|S|}$ is the depot, and the depot does not appear in any other operation.
Moreover, for each $\ell \in \{1, \dots, {|S|} - 1\}$, the last node of~$r_\ell$ equals the 
first node of $r_{\ell + 1}$.
The total time of the solution, denoted by $t(S)$, is simply the sum of times of the operations,
that is, $t(S) = \sum_{\ell=1}^{|S|} t(o_\ell)$.
The goal of the FSTSP is to find a solution with minimum time.

It is worth mentioning that the definition given by Murray and Chu~\cite{MurrayC2015} differs
from ours in three aspects: 
($i$) there is a flight endurance for the drone, which limits the nodes it can reach; 
($ii$) some nodes are exclusive to the truck (due to package size, for instance);
($iii$) there are times required for the truck driver to prepare the drone for launch and 
for the drone to recover after delivery.
We decided to remove these features because the benchmark instances~\cite{BoumanAS2018}
used by other works with which we are comparing ours also disregard them.

As already mentioned, the order-first split-second heuristics are a very common approach
to handle the FSTSP.
We will thus formally define the \defi{h-FSTSP}: given the same input as the FSTSP 
($n, N, t_R, t_D$) plus a Hamiltonian cycle~$h$ over~$N = \{0, 1, \ldots, n+1\}$, one wants
to find a solution for the FSTSP of minimum cost that \defi{respects} the order of sequence~$h$, 
meaning that the nodes visited by the truck follow the order in which they appear in~$h$ 
and that the drone always visits nodes that are between the launch and rendezvous nodes.
The next result establishes an important relationship between these two problems.

\begin{lemma}
\label{lem:relation_fstsp_hfstsp}
    Given an optimal solution~$S^*$ for the FSTSP over instance ($n, N, t_R, t_D$), there
    exists a Hamiltonian cycle~$h$ over~$N$ such that~$S^*$ respects~$h$.
\end{lemma}
\begin{proof}
    Let $S^* = ((r_1,d_1), \ldots, (r_{|S^*|}, d_{|S^*|}))$ be an optimal solution for the FSTSP.
    For each $r_\ell = (v^{r_\ell}_1, \ldots, v^{r_\ell}_{|r_\ell|})$ such that 
    $d_\ell = (v^{r_\ell}_1, v^{d_\ell}, v^{r_\ell}_{|r_\ell|})$,
    let $r'_\ell = (v^{r_\ell}_1, v^{d_\ell}, v^{r_\ell}_2, \ldots, v^{r_\ell}_{|r_\ell|})$.
    For each $r_\ell$ such that $d_\ell = ( )$, let $r'_\ell = r_\ell$.
    At last, build~$h$ by concatenating the sequences $r'_\ell$, for each $\ell = 1, \ldots, |S^*|$.
    By construction, $S^*$ respects~$h$.
\end{proof}

%%%%%%%%%%%%%%%%%%%%%%%%%%%%%%%%%%%%%%%%%%%%%%%%%%%%%%%%%%%%%%%%%%%%%%%%%%%%%%%
\section{The split algorithm}
\label{sec:split}

Algorithm~\ref{alg:split_trad} is a version of the Split Algorithm, proposed by
Ha \textit{et al.}~\cite{HaDPH2018}, with a more careful implementation to avoid time
complexity~$O(n^4)$.
In fact, both its best and worst cases have~$\Theta(n^3)$ efficiency.

\begin{algorithm}[!htb]
\small
  \sffamily
  \begin{algorithmic}[1]
    \Require{a Hamiltonian cycle~$h = (v_0,v_1,\ldots,v_{n+1})$ over $N$, $t_R(u, v)$ and $t_D(u, v)$ for any pair $u, v \in N$}
    \Ensure{a solution $S$ for the FSTSP}
    
    \State let $G$ be a digraph with $V(G) = N$ and $E(G) = \emptyset$ \CommentHigh{Each arc will correspond to an operation}
    \For{$i = 0$ to $n$} \label{line:split:init_G}
    \CommentHigh{Initially, each operation is not using the drone}
        \State add arc $v_i v_{i+1}$ to $E(G)$ with cost $t_R(v_i, v_{i+1})$ and no drone node associated ($-1$)
    \EndFor
    \For{$i = 0$ to $n - 1$} \label{line:split:launch}
        \For{$k = i + 2$ to $n + 1$} \label{line:split:rendesvouz}
            \State $\var{truck\_cost} \gets \sum_{\ell = i}^{k - 1} t_R(v_\ell, v_{\ell+1})$ \label{line:split:truck_cost} \CommentHigh{Important to avoid complexity $O(n^4)$}
            \State $\var{drone\_node} \gets -1$
            \State $\var{min\_cost} \gets \var{truck\_cost}$
            \For{$j = i + 1$ to $k - 1$} \CommentHigh{Find a node to serve with drone instead of truck} \label{line:split:drone_node}
                \State $\var{drone\_cost} \gets t_D(v_i, v_j) + t_D(v_j, v_k)$
                \State $\var{delta} \gets t_R(v_{j-1}, v_{j+1}) - \big(t_R(v_{j-1}, v_j) + t_R(v_j, v_{j+1})\big)$ \label{line:split:cost_op1}
                \State $\var{new\_cost} \gets \max\{\var{drone\_cost}, \, \var{truck\_cost} + \var{delta}\}$ \label{line:split:cost_op2}
                \If{$\var{new\_cost} < \var{min\_cost}$}
                    \State $\var{min\_cost} \gets \var{new\_cost}$
                    \State $\var{drone\_node} \gets v_j$
                \EndIf
            \EndFor
            \State add arc $v_i v_k$ to $E(G)$ with $\var{min\_cost}$ and associated with $\var{drone\_node}$ \label{line:split:best_operation}
        \EndFor
    \EndFor
    \State Find a shortest path $P$ from $v_0$ to $v_{n+1}$ in $G$ \label{line:split:shortest_path}
    \State Build a solution $S$ with the operations corresponding to each arc in $P$ \label{line:split:solution}
    \State \Return $S$
  \end{algorithmic}
  \caption{\textsc{SplitAlgorithm}($h$, $N$, $n$, $t_R$, $t_D$)}
  \label{alg:split_trad}
\end{algorithm}

We consider a Hamiltonian cycle~$h$ over set $N = \{0,\ldots,n+1\}$ as any sequence 
$(v_0,v_1,\ldots,v_{n+1})$ such that both $v_0 = 0$ and $v_{n+1} = n+1$ correspond to the depot
and every $v_i \in \{1, \ldots, n\}$, with $1 \leq i \leq n$.
The algorithm builds a digraph~$G$ with vertex set~$N$ in which each arc $v_i v_k \in E(G)$
corresponds to an operation and, thus, can be associated with a drone node.
If no drone node is associated with the arc, then we represent this with~$-1$ in the 
pseudocode, and the truck path of the operation is simply $r = (v_i, v_{i+1}, \ldots, v_k)$
while the drone path is empty.
If a drone node~$v_j$ is associated with arc $v_i v_k$, then $i < j < k$ and the truck path
is $r = (v_i, v_{i+1}, \ldots, v_{j-1}, v_{j+1}, \ldots, v_k)$ while the drone path is
$d = (v_i, v_j, v_k)$.
We consider the cost of each arc as being the time of the operation. %truck path, that is, $t(r)$. \carla{?? the time of the operation, right?}

Initially, the algorithm adds arcs to~$G$ which simply correspond to the path which 
follows~$h$ (line~\ref{line:split:init_G}).
At this point, therefore, they describe a solution where the drone is never used, 
and the truck visits all clients by itself.

The core of the algorithm is composed of three nested loops.
With the first and second loops (lines~\ref{line:split:launch} and~\ref{line:split:rendesvouz}),
the algorithm considers each possible pair of nodes $(v_i, v_k)$ to be the launch 
and rendezvous nodes, respectively. 
Initially, the truck path is $r = (v_i, \ldots, v_k)$ and its cost is calculated in line~\ref{line:split:truck_cost} ($\var{truck\_cost} \gets t(r)$), and no drone node 
is associated with the operation ($\var{drone\_node} \gets -1$).
In the third nested loop (line~\ref{line:split:drone_node}), the algorithm verifies every 
drone node~$v_j$, with~$j$ between~$i$ and~$k$, to discover which drone node gives the 
operation of smallest cost (line~\ref{line:split:best_operation}).
We refer to the sequence $(i,j,k)$ as a \defi{triple} being considered by the algorithm.
We improved the time complexity of the algorithm presented by Ha \textit{et al.}~\cite{HaDPH2018} 
by carefully computing the cost of each possible operation in constant time 
(lines~\ref{line:split:cost_op1} and~\ref{line:split:cost_op2}).

Once~$G$ is built, the algorithm finds a shortest path in it and uses this path to build a
solution for the FSTSP (lines~\ref{line:split:shortest_path} and~\ref{line:split:solution}),
which is an optimal solution for the h-FSTSP considering~$h$~\cite{AgatzBS2018, HaDPH2018}.
Recall that this involves creating, for each arc $v_i v_k$ in the shortest path, an operation
$(r,d)$ for which the drone is used or not, according to the drone node associated with the arc,
as mentioned before.
This along with Lemma~\ref{lem:relation_fstsp_hfstsp} shows that it is always possible to find an optimal solution for the FSTSP using Algorithm~\ref{alg:split_trad} with an appropriate~$h$.

%%%%%%%%%%%%%%%%%%%%%%%%%%%%%%%%%%%%%%%%%%%%%%%%%%%%%%%%%%%%%%%%%%%%%%%%%%%%%%%
\section{The lazy drone property}
\label{sec:lazy_property}

%\url{https://trello.com/c/rfnGebYK/23-lazy-drone-property}

In this section, consider a fixed sequence $h = (v_1,\ldots,v_{|h|})$ of nodes.
Also, for any three nodes $v_i, v_j, v_k$ such that $1 \leq i < j < k \leq |h|$, we define  
the operation $o_{i,j,k} = (r,d)$ where 
$r = (v_i, \ldots, v_{j-1}, v_{j+1}, \ldots, v_k)$ and $d = (v_i, v_j, v_k)$.
We thus have 
\begin{equation*}
    t(o_{i,j,k}) = \max\left\{ 
        \sum_{\ell=i}^{j-2} t_R(v_\ell, v_{\ell+1}) + t_R(v_{j-1}, v_{j+1}) + \sum_{\ell=j+1}^{k-1} t_R(v_\ell, v_{\ell+1}),
        \,\, 
        t_D(v_i, v_j) + t_D(v_j, v_k)
    \right\} \, ,
\end{equation*}
where the first term is $t(r)$ and the second is $t(d)$.

For $i < j < k$ and $i' < j' < k'$, we say that operation \defi{$o_{i,j,k}$ is contained in operation $o_{i',j',k'}$ with respect to sequence~$h$} if $i' \leq i$ and $k' \geq k$. 
Moreover, we say \defi{$o_{i,j,k}$ dominates $o_{i',j',k'}$ with respect to~$h$} if
the former operation is contained in the latter
and if 
\begin{equation*}
    t(o_{i',j',k'}) \geq \sum_{\ell=i'}^{i-1} t_R(v_\ell, v_{\ell+1}) + t(o_{i,j,k}) + \sum_{\ell=k}^{k'-1} t_R(v_\ell, v_{\ell+1}) \enspace.
\end{equation*}
Note that both sides of this inequality correspond to the costs of visiting the same set of nodes,
and that they both start and end at the same nodes.
The following result shows that operations that are dominated by others are not necessary to achieve an optimal solution.

\begin{lemma} 
\label{lem:opt_dom}
    There exists an optimal solution for the FSTSP (or h-FSTSP) that does not have dominated operations.
\end{lemma}
\begin{proof}
    Consider an optimal solution~$S^*$ for the FSTSP with an operation $o_{i',j',k'} = (r',d')$,
    which is dominated by $o_{i,j,k} = (r,d)$ with respect to some sequence $h = (v_1, \dots, v_{|h|})$.
    Consider another solution~$S$ build from~$S^*$ which replaces $o_{i',j',k'}$ by the three 
    operations $((v_{i'}, \ldots, v_i), ())$, $(r,d)$, and $((v_k, \ldots, v_{k'}), ())$.
    Now note that
    \begin{equation*}
        t(S^*) - t(S) = t(o_{i',j',k'}) - \left( \sum_{\ell=i'}^{i-1} t_R(v_\ell v_{\ell+1}) + t(o_{i,j,k}) + \sum_{\ell=k}^{k'-1} t_R(v_\ell v_{\ell+1}) \right) \geq 0 \enspace ,
    \end{equation*}
    where the last inequality holds because $o_{i,j,k}$ dominates $o_{i',j',k'}$ with respect to~$h$.
\end{proof}

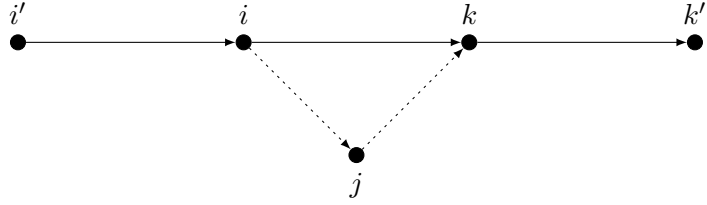
\begin{figure}[!htb]
\centering
\begin{tikzpicture}[scale=1.5]
    \node[draw,fill,circle,inner sep=2pt,label=above:$i'$] (i') at (0,0) {};
    \node[draw,fill,circle,inner sep=2pt,label=above:$i$] (i) at (2,0) {};
    \node[draw,fill,circle,inner sep=2pt,label=above:$k$] (k) at (4,0) {};
    \node[draw,fill,circle,inner sep=2pt,label=above:$k'$] (k') at (6,0) {};
    \node[draw,fill,circle,inner sep=2pt,label=below:$j$] (j) at (3,-1) {};

    \draw[-latex] (i') -- (i);
    \draw[-latex] (i) -- (k);
    \draw[-latex] (k) -- (k');

    \begin{scope}[dashed,dash pattern=on 1pt off 2pt] 
        \draw[-latex] (i) -- (j);
        \draw[-latex] (j) -- (k);
    \end{scope}
\end{tikzpicture}
\caption{If the drone is fast in operation~$o_{i,j,k}$, then there is no advantage in~$o_{i',j,k'}$. \label{fig:lazyprop}}
\end{figure}

We say the drone is \defi{fast in operation $o_{i,j,k}$} if $t(d) \leq t(r)$, that is, 
the drone does not leave the truck waiting.
This implies that $t(o_{i,j,k}) = t(r)$.
The following result states that if the drone is fast in operation $o_{i,j,k}$, then, 
for any $i' \leq i$ and $k' \geq k$, replacing $o_{i,j,k}$ by $o_{i',j,k'}$ in a solution,
i.e., sending the drone from node~$v_{i'}$ to serve~$v_j$ and return to node~$v_{k'}$ 
instead of sending it from~$v_i$ to~$v_j$ and then~$v_k$, cannot decrease the cost of the solution. 
Figure~\ref{fig:lazyprop} illustrates such scenario.

\begin{lemma}
\label{lem:fast_dom}
    Let $h = (v_1,\ldots,v_{|h|})$ be a sequence of nodes and let $v_i, v_j, v_k$ be three nodes such that
    $1 \leq i < j < k \leq |h|$.
    If the drone is fast in operation $o_{i,j,k}$, then for any pair of nodes $v_{i'},v_{k'}$ with $i' \leq i$
    and $k' \geq k$, $o_{i,j,k}$ dominates $o_{i',j,k'}$ with respect to $h$.
\end{lemma}
\begin{proof}
    By definition, 
    \begin{equation*}
        t(o_{i',j,k'}) = \max\left\{ 
        \sum_{\ell=i'}^{j-2} t_R(v_\ell v_{\ell+1}) + t_R(v_{j-1}, v_{j+1}) + \sum_{\ell=j+1}^{k'-1} t_R(v_\ell, v_{\ell+1}),
        \,\, 
        t_D(v_{i'}, v_j) + t_D(v_j, v_{k'})
        \right\} \enspace .
    \end{equation*}
    Thus, clearly, 
    \begin{equation}
    \label{eq:time_newop}
        t(o_{i',j,k'}) \geq 
        \sum_{\ell=i'}^{j-2} t_R(v_\ell, v_{\ell+1}) + t_R(v_{j-1}, v_{j+1}) + \sum_{\ell=j+1}^{k'-1} t_R(v_\ell, v_{\ell+1}) \enspace .
    \end{equation}
    
    Since the drone is fast in operation $o_{i,j,k}$, we know that
    \begin{equation*}
        t(o_{i,j,k}) = \sum_{\ell=i}^{j-2} t_R(v_\ell, v_{\ell+1}) + t_R(v_{j-1}, v_{j+1}) + \sum_{\ell=j+1}^{k-1} t_R(v_\ell, v_{\ell+1}) \enspace ,
    \end{equation*}
    which, together with Eq.~\eqref{eq:time_newop}, and because $i' \leq i$ and $k' \geq k$, means that
    \begin{equation*}
        t(o_{i',j,k'}) \geq \sum_{\ell=i'}^{i-1} t_R(v_\ell, v_{\ell+1}) + t(o_{i,j,k}) + \sum_{\ell=k}^{k'-1} t_R(v_\ell, v_{\ell+1}) \enspace .
    \end{equation*}
    Thus, by definition, $o_{i,j,k}$ dominates $o_{i',j,k'}$.
\end{proof}

The previous result establishes a condition to identify a family of dominated operations 
from a given operation~$o_{i,j,k}$. 
The following result shows how this idea may help an algorithm for the h-FSTSP 
to speed up by ignoring unnecessary operations.

\begin{theorem}[Lazy Drone Property]
    Let $h = (v_1,\ldots,v_{|h|})$ be a sequence of nodes and let $v_i, v_j, v_k$ 
    be three nodes such that $1 \leq i < j < k \leq |h|$.
    If the drone is fast in operation $o_{i,j,k}$, then any other operation $o_{i',j,k'}$ 
    with $i' \leq i$ and $k' \geq k$ can be ignored by an algorithm for the h-FSTSP.
\end{theorem}
\begin{proof}
    Follows directly from Lemmas~\ref{lem:opt_dom} and~\ref{lem:fast_dom}.
\end{proof}

Thus, we conclude that a fast drone can be lazy in the sense that it does not need to consider
longer flights once it finds an operation that does not leave the truck waiting. 
Furthermore, this allows an order-first split-second algorithm to be lazy with respect 
to the operations it considers with no risk of losing the optimum.

%%%%%%%%%%%%%%%%%%%%%%%%%%%%%%%%%%%%%%%%%%%%%%%%%%%%%%%%%%%%%%%%%%%%%%%%%%%%%%%
\section{The split lazy algorithm}
\label{sec:split_lazy}

%\url{https://trello.com/c/SqWwgxfB/40-algoritmo-split-lazy-para-tsp-d}

In this section, we present the Split Lazy Algorithm in Algorithm~\ref{alg:split_lazy}, which uses the Lazy Drone Property to optimally solve the h-FSTPS much faster in practice 
(almost linear time), despite not reducing the worst-case time complexity from~$O(n^3)$.

\begin{algorithm}[!htb]
\small
  \sffamily
  \begin{algorithmic}[1]
    \Require{a Hamiltonian cycle~$h = (v_0,v_1,\ldots,v_{n+1})$ over $N$, $t_R(u, v)$ and $t_D(u, v)$ for any pair $u, v \in N$}
    \Ensure{a solution $S$ for the FSTSP}

    \State let $G$ be a digraph with $V(G) = N$ and $E(G) = \emptyset$ \CommentHigh{Each arc will correspond to an operation}
    \For{$i = 0$ to $n$} \label{line:splitlazy:init_G}
    \CommentHigh{Initially, each operation is not using the drone}
        \State add arc $v_i v_{i+1}$ to $E(G)$ with cost $t_R(v_i, v_{i+1})$ and no drone node associated
    \EndFor
    \For{$j = 1$ to $n$} \label{line:splitlazy:drone_node} \CommentHigh{consider each drone node $v_j$}
        \State $\var{base\_cost} \gets t_R(v_{j-1}, v_{j+1})$ \label{line:splitlazy:base_cost}
        \State $\var{pre\_cost} \gets - t_R(v_{j-1}, v_{j})$ \label{line:splitlazy:pre_cost1}
        \State $k_{max} \gets n+1$
        \For{$i = j-1$ down to $0$} \label{line:splitlazy:launch} \CommentHigh{consider each launch node $v_i$}
            \State $k \gets j + 1$ \label{line:splitlazy:rendezvous1} \CommentHigh{consider smallest rendezvous node $v_k$}
            \State $\var{drone\_cost} \gets t_D(v_{i}, v_{j}) + t_D(v_{j}, v_{k})$
            \State $\var{pre\_cost} \gets \var{pre\_cost} + t_R(v_{i}, v_{i+1})$ \label{line:splitlazy:pre_cost2}
            \State $\var{pos\_cost} \gets 0$ \label{line:splitlazy:pos_cost1}
            \State $\var{truck\_cost} \gets \var{pre\_cost} + \var{base\_cost} + \var{pos\_cost}$ \label{line:splitlazy:truck_cost1}
            \If{arc $v_i v_k \not \in E(G)$ or its cost is greater than $\max \{\var{drone\_cost}, \var{truck\_cost}\}$}
                 \State add or update arc $v_i v_k$ with cost $\max\{\var{drone\_cost}, \var{truck\_cost}\}$ and drone node $v_j$ \label{line:splitlazy:update1}
            \EndIf
            \If{$\var{drone\_cost} \leq \var{truck\_cost}$} \label{line:splitlazy:fast_op1} \CommentHigh{found a fast operation $o_{i,j,k}$}
                \State \textbf{break}
            \EndIf
            \For{$k = j+2$ to $k_{max}$} \label{line:splitlazy:rendezvous2} \CommentHigh{consider each rendezvous node~$v_k$ with $k \geq j + 2$}
                \State $\var{drone\_cost} \gets t_D(v_{i}, v_{j}) + t_D(v_{j}, v_{k})$
                \State $\var{pos\_cost} \gets \var{pos\_cost} + t_R(v_{k-1}, v_{k})$ \label{line:splitlazy:pos_cost2}
                \State $\var{truck\_cost} \gets \var{pre\_cost} + \var{base\_cost} + \var{pos\_cost}$ \label{line:splitlazy:truck_cost2}
                \If{arc $v_i v_k \not \in E(G)$ or its cost is greater than $\max \{\var{drone\_cost}, \var{truck\_cost}\}$}
                    \State add or update $v_i v_k$ with cost $\max \{\var{drone\_cost}, \var{truck\_cost}\}$ and drone node $v_j$ \label{line:splitlazy:update2}
                \EndIf
                \If{$\var{drone\_cost} \leq \var{truck\_cost}$} \label{line:splitlazy:fast_op2} \CommentHigh{found a fast operation $o_{i,j,k}$}
                    \State $k_{max} \gets k-1$ 
                    \State \textbf{break}
                \EndIf
            \EndFor
        \EndFor
    \EndFor
    \State Find a shortest path $P$ from $v_0$ to $v_{n+1}$ in $G$ \label{line:splitlazy:shortest_path}
    \State Build a solution $S$ with the operations corresponding to each arc in $P$ \label{line:splitlazy:solution}
    \State \Return $S$
  \end{algorithmic}
  \caption{\textsc{SplitLazyAlgorithm}($H$, $N$, $n$, $t_R$, $t_D$)}
  \label{alg:split_lazy}
\end{algorithm}

This algorithm has two similarities with the Split Algorithm presented in 
Section~\ref{sec:split}.
First, we also consider a initial Hamiltonian cycle~$h = (v_0,v_1,\ldots,v_{n+1})$ over 
set $N = \{0,\ldots,n+1\}$ where $v_0 = 0$ and $v_{n+1} = n+1$ correspond to the depot 
and every $v_i \in \{1,\ldots,n\}$, with $1 \leq i \leq n$.
Second, it also builds a digraph~$G$ with vertex set~$N$ in which each arc $v_i v_k$
corresponds to an operation and can be associated with a drone node.
The truck path and node path of such operation is thus built exactly as described in 
Section~\ref{sec:split}.
Thus, initially, it also adds arcs to~$G$ which correspond to the path which follows~$h$
(line~\ref{line:splitlazy:init_G}).

The core of the algorithm is also composed of three nested loops, used to test the possible triples $(i,j,k)$.
However, in contrast with Algorithm~\ref{alg:split_trad}, the first loop considers each 
node~$v_j$ candidate to be a drone node in some operation (line~\ref{line:splitlazy:drone_node}).
In the second and third nested loops (lines~\ref{line:splitlazy:launch} and~\ref{line:splitlazy:rendezvous2}), the algorithm considers pairs $(v_i, v_k)$ 
of possible launch and rendezvous nodes, respectively, which gradually move away from~$v_j$.
Specifically, index~$i$ decreases from~$j-1$ to~$0$, while index~$k$ increases from~$j+1$ to~$n+1$. 
If the algorithm detects that the drone is fast in operation $o_{i,j,k}$, in some iteration 
of the second or third loops, it gives up trying smaller indices for the launch or larger 
indices for the rendezvous nodes, respectively (lines~\ref{line:splitlazy:fast_op1} 
and~\ref{line:splitlazy:fast_op2}).
It can do that because operation $o_{i,j,k}$ dominates the operations which will not be 
considered.

Let $o_{i,j,k} = (r,d)$ be an operation being considered by the algorithm, which means
that $r = (v_i, \ldots, v_{j-1}, v_{j+1}, \ldots, v_k)$ and $d = (v_i, v_j, v_k)$.
To compute the cost~$t(r)$ of the truck path, the algorithm only needs to sum 
$\var{base\_cost}$, $\var{pre\_cost}$, and $\var{pos\_cost}$ 
(lines~\ref{line:splitlazy:truck_cost1} and~\ref{line:splitlazy:truck_cost2}).
The $\var{base\_cost}$ is simply the cost for the truck to go straight from~$v_{j-1}$ 
to~$v_{j+1}$ (line~\ref{line:splitlazy:base_cost}).
The $\var{pre\_cost}$ is the cost of the truck from~$v_i$ to~$v_{j-1}$ following~$h$ 
(lines~\ref{line:splitlazy:pre_cost1} and~\ref{line:splitlazy:pre_cost2}), 
while the $\var{pos\_cost}$ is the cost of the truck from~$v_{j+1}$ to~$v_k$ 
(lines~\ref{line:splitlazy:pos_cost1} and~\ref{line:splitlazy:pos_cost2}).
Notice that the value initially attributed to $\var{pre\_cost}$ is canceled by 
its update in the first iteration of the second nested loop.
Splitting the truck cost in this way is important because it allows us to update it 
in constant time simply when~$i$ is decremented or~$k$ is incremented.
In particular, when~$i$ changes, the algorithm has to reset $\var{pos\_cost}$ 
(line~\ref{line:splitlazy:pos_cost1}), but the $\var{pre\_cost}$ is simply incremented.

Notice that it is important to separately consider the rendezvous index~$j+1$ before 
the third nested loop (line~\ref{line:splitlazy:rendezvous1}) because if the drone is fast
in operation $o_{i,j,j+1}$, then this operation dominates any other operation with drone 
node~$v_j$ and launch node with an index smaller than~$i$.
Thus, it allows us to jump to the next drone node~$j+1$ (line~\ref{line:splitlazy:fast_op1}). 
In contrast, a fast operation $o_{i,j,k}$ with $k > j+1$ does not dominate operations with 
launch node index smaller than~$i$ and rendezvous node index smaller than~$k$, i.e.,
in this case, we still need to consider drone node~$v_j$ with launch nodes whose indices 
are smaller than~$i$.
Moreover, $k_{max}$ is the index of the farthest rendezvous node in~$h$ that may have 
an operation around drone node~$v_j$ that is not dominated by some operation already 
considered by the algorithm.
In particular, $k_{max}$ is updated inside the third nested loop when $o_{i,j,k}$ is fast,
because any operation $o_{i',j,k'}$ with $i' \leq i$ and $k' \geq k$ is dominated 
by~$o_{i, j, k}$ (line~\ref{line:splitlazy:fast_op2}).

Let~$v_j$ be a drone node that is being considered by the algorithm and assume that it
finds an operation $o_{i,j,k}$ that causes it to add to $E(G)$ the arc $v_i v_k$ associated
with drone node~$v_j$.
Now, note that in a later iteration, when considering a drone node~$v_{j'}$ 
with index $j' > j$, the algorithm may find an operation $o_{i,j',k}$ with cost smaller
than that of $o_{i,j,k}$.
Since both operations are represented by the same arc $v_i v_k$, the algorithm has to update
the cost and the drone node of such arc.
That is why Algorithm~\ref{alg:split_lazy} has to be able to update arcs instead of simply 
adding them, or dealing with multiple arcs (lines~\ref{line:splitlazy:update1} 
and~\ref{line:splitlazy:update2}).

At last, once~$G$ is built, the algorithm moves on in the same way as 
Algorithm~\ref{alg:split_trad}: it finds a shortest path in~$G$ and it uses this path 
to build a solution for the FSTSP (lines~\ref{line:splitlazy:shortest_path} and~\ref{line:splitlazy:solution}).
This solution is optimal for the h-FSTSP considering~$h$, despite the fact that 
Algorithm~\ref{alg:split_lazy} does not necessarily consider all $n \choose 3$ 
operations to build~$G$.
This happens because the operations not considered are dominated by some operation 
considered.
This, along with Lemma~\ref{lem:relation_fstsp_hfstsp}, shows that it is always possible
to find an optimal solution for the FSTSP using Algorithm~\ref{alg:split_lazy} 
with an appropriate~$h$.

%%%%%%%%%%%%%%%%%%%%%%%%%%%%%%%%%%%%%%%%%%%%%%%%%%%%%%%%%%%%%%%%%%%%%%%%%%%%%%%
\section{Computational study}
\label{sec:computational}

To test our Split Lazy Algorithm, we need benchmark instances for the FSTSP, heuristics to generate initial Hamiltonian cycles for each instance, and other algorithms for the h-FSTSP with which to compare.
In the next three sections, we describe our options for these important choices.
In Section~\ref{sub:sec:lazy_algorithms} we describe some implementation choices for our
Split Lazy Algorithm, which resulted in two final versions.
At last, in Section~\ref{sub:sec:results}, we show that, for the benchmark instances
chosen, the number of non-dominated operations is linear in~$n$, which allows for 
a great improvement in time efficiency.

\subsection{FSTSP instances}
\label{sub:sec:instances}
The benchmark instances we use are the ones by Bouman \textit{et al.}~\cite{BoumanAS2018}.
The graphs from this benchmark are all planar and the Euclidean distance is considered.
They are divided into three types.
In the \defi{uniform} one, the $(x,y)$ coordinates of a vertex are drawn independently 
and uniformly at random from $\{0,1,\ldots,100\}$.
In the \defi{1-center} one, the coordinates of a vertex are $(r \cos a, r \sin a)$, 
where~$a$ is an angle drawn from $[0,2\pi]$ uniformly and~$r$ is a distance drawn from a normal
distribution with mean~0 and standard deviation~50.
For the \defi{2-center}, the difference from the 1-center is that every location is translated
by~200 units over the $x$ axis with probability~$\frac12$.
Also, for the three types of instances, there are three types of drone velocity~$\alpha$,
which are 1, 2, or 3, and mean that the drone and the truck have equal speed, the drone is 
twice as fast, or the drone is three times as fast, respectively.
At last, the amount of vertices in the instances varies in $\{5,6,7,8,9,10,20,50,75,100,175,250,375,500\}$.

\subsection{Hamiltonian cycles}
\label{sub:sec:hamiltonian_cycles}
Our main goal is to test the performance of the Split Lazy Algorithm, which means
that we need at least an initial Hamiltonian cycle for each instance.
To strengthen the result, we have chosen to do it with three different initial cycles.
In particular, we chose three classical heuristics for building feasible solutions 
for the TSP.

The first one is the Lin-Kernighan heuristic~\cite{LinK1973} (LKH)\footnote{Available
at \url{https://github.com/RodolfoPichardo/LinKernighanTSP}.}, which has been shown 
to perform really well and to be one of the most effective methods for the TSP~\cite{Helsgaun2000}.
The second one is the Christofides algorithm~\cite{Christofides1976,Christofides2022} (CHR), 
which was devised specifically for weight functions that respect the triangular inequality,
but can be used to build feasible solutions for any weight function.
The last one is the Nearest Neighbor heuristic~\cite{BellmoreN1968} (NNH), which was one of the 
first heuristics for the TSP and it is a very simple and efficient greedy algorithm.

\subsection{Existing algorithms for h-FSTSP}
\label{sub:sec:fstsp_heuristics}
We have chosen to compare the Split Lazy Algorithm with three different existing
algorithms for the h-FSTSP.
The first one is the Split Algorithm by Ha \textit{et al.}~\cite{HaDPH2018},
the second one is the dynamic programming algorithm by Agatz \textit{et al.}~\cite{AgatzBS2018}, and
the third one is the algorithm by Kundu \textit{et al.}~\cite{KunduEM2022}.
We will call them, respectively, \textsc{Split}, \textsc{Agatz}, and \textsc{Kundu}.

We used our own implementation of \textsc{Split}, which follows what was described 
in the original paper and represents the graph by an adjacency matrix.
The code for \textsc{Agatz} was made available online by the authors\footnote{Available 
at \url{https://github.com/pcbouman-eur/Drones-TSP}.}.
Although the code for \textsc{Kundu} was not made available by the authors, we found it 
at Kundu's thesis~\cite{Kundu2022}.
It is worth mentioning that the algorithm described in the paper~\cite{KunduEM2022} differs
from the implementation in the sense that the implementation allows some ``breaks'' in 
the loops that are not explained in the algorithm.
In a sense, they look similar to some of the decisions we do because of the Lazy Drone Property.
However, they are not considered by the authors, which explain the better performance of their
algorithm, with respect to \textsc{Agatz}, due to other features.

\subsection{Split lazy algorithms}
\label{sub:sec:lazy_algorithms}

We implemented several versions of the Split Lazy Algorithm, according to different ways of
representing the graph.
The first and simpler implementation, called \textsc{LazyMatriz}, used an adjacency matrix. 
Other implementations used an adjacency list, implemented with dynamic vectors or linked lists.
Moreover, since the same arc of the graph may correspond to different triples (with 
the same origin and destiny, but different drone nodes), we either had to implement an
update arc operation, or use a multigraph.
The best version was the combination of linked lists and multigraph, which we call 
\textsc{LazyLists}.

\subsection{Empirical results}
\label{sub:sec:results}

%\textcolor{blue}{Tentar colocar a partir daqui uma estrutura para a parte de análise dos resultados. Antes, os links das planilhas pra pegar os dados.}

%\url{https://docs.google.com/spreadsheets/d/1z_RP_ehpSaQ9WmYURFWhfZFFhtdzsfkkA09asg62XYk/edit?usp=sharing}

%\url{https://docs.google.com/spreadsheets/d/1-zGMsT9mfJHAJ2jMVk6Bhbct_UuJCBa3uSouezIxS50/edit?usp=sharing}

%\url{https://docs.google.com/spreadsheets/d/1xvuluqyI4zwRfI-35iIgBDGyyBfWW8f5wr1wmCwMIik/edit?usp=sharing}

The results obtained from using the Hamiltonian cycles produced by LKH, CHR, and NNH exhibited similar patterns. 
For this reason, we focus on the LKH and provide a summarized overview of the other two.

Table~\ref{tab:times_lkh} shows each algorithm's average time
to solve the h-FSTSP considering a Hamiltonian cycle produced by LKH.
Each row in the table corresponds to a benchmark instance size (except for the first row,
which merges instances with sizes between 5 and 10). 
It is worth noting that all tested algorithms obtain an optimal solution for the h-FSTSP. 
The cost reduction is computed for each instance and uses the formula (TSP cost $-$ h-FSTSP cost) / TSP cost, where TSP cost is the cost of the initial Hamiltonian cycle and h-FSTSP cost is obtained by the tested algorithms.
The average reductions in solution costs are between 20\% and 30\%, with the reduction value slightly decreasing as instance sizes grow. 

\begin{table}[!htb]
    \centering
    \caption{Comparison among algorithms solving h-FSTSP from a cycle produced by LKH.}
    \label{tab:times_lkh}
    \resizebox{0.8\textwidth}{!}{%
    \begin{tabular}{|c|c|c|c|c|c|c|c|}
    \hline
    Number of & Number of & Cost & \textsc{Agatz} & \textsc{Split} & \textsc{Kundu} & \textsc{LazyMatrix} & \textsc{LazyLists} \\
    Nodes & Instances & Reduction & Time (ns) & Time (ns) & Time (ns) & Time (ns) & Time (ns) \\ \hline\hline
    $\leq$ 10 & 540 & -26.90\% & 4,527,379 & 58,906 & 14,887,616 & 31,839 & 25,255 \\ \hline
    20 & 90 & -23.44\% & 5,694,050 & 552,617 & 15,021,452 & 59,599 & 37,421 \\ \hline
    50 & 90 & -25.48\% & 12,811,271 & 7,738,033 & 15,908,985 & 149,856 & 74,877 \\ \hline
    75 & 90 & -24.96\% & 24,868,307 & 26,071,367 & 16,610,260 & 244,693 & 97,099 \\ \hline
    100 & 90 & -23.80\% & 33,659,476 & 61,769,128 & 19,514,029 & 382,266 & 121,584 \\ \hline
    175 & 90 & -23.53\% & 73,915,942 & 326,161,167 & 22,609,606 & 900,591 & 191,275 \\ \hline
    250 & 180 & -23.10\% & 139,289,609 & 952,474,533 & 27,488,362 & 1,648,081 & 262,040 \\ \hline
    375 & 127 & -23.13\% & 363,827,911 & 3,227,132,992 & 35,247,873 & 3,555,986 & 393,535 \\ \hline
    500 & 92 & -22.34\% & 799,852,512 & 7,659,839,022 & 43,484,661 & 6,134,871 & 514,417 \\ \hline
    \end{tabular}
    }
\end{table}

Despite all algorithms achieving the same solution cost, their average running times differ significantly.
Focusing on the largest instances, which have 500 nodes each, we notice that \textsc{Split} 
is the slowest, taking 7.66 seconds, \textsc{Agatz} takes 0.80 seconds, \textsc{Kundu} takes 43 
milliseconds, \textsc{LazyMatrix} takes 6.1 milliseconds, and \textsc{LazyLists} takes 0.51 
milliseconds, on average.
Thus, for such instances, our methods are, respectively, 7 and 84 times faster than the best literature results that we know of.

\begin{figure}[!htb]
    \centering
    \includegraphics[width=\textwidth]{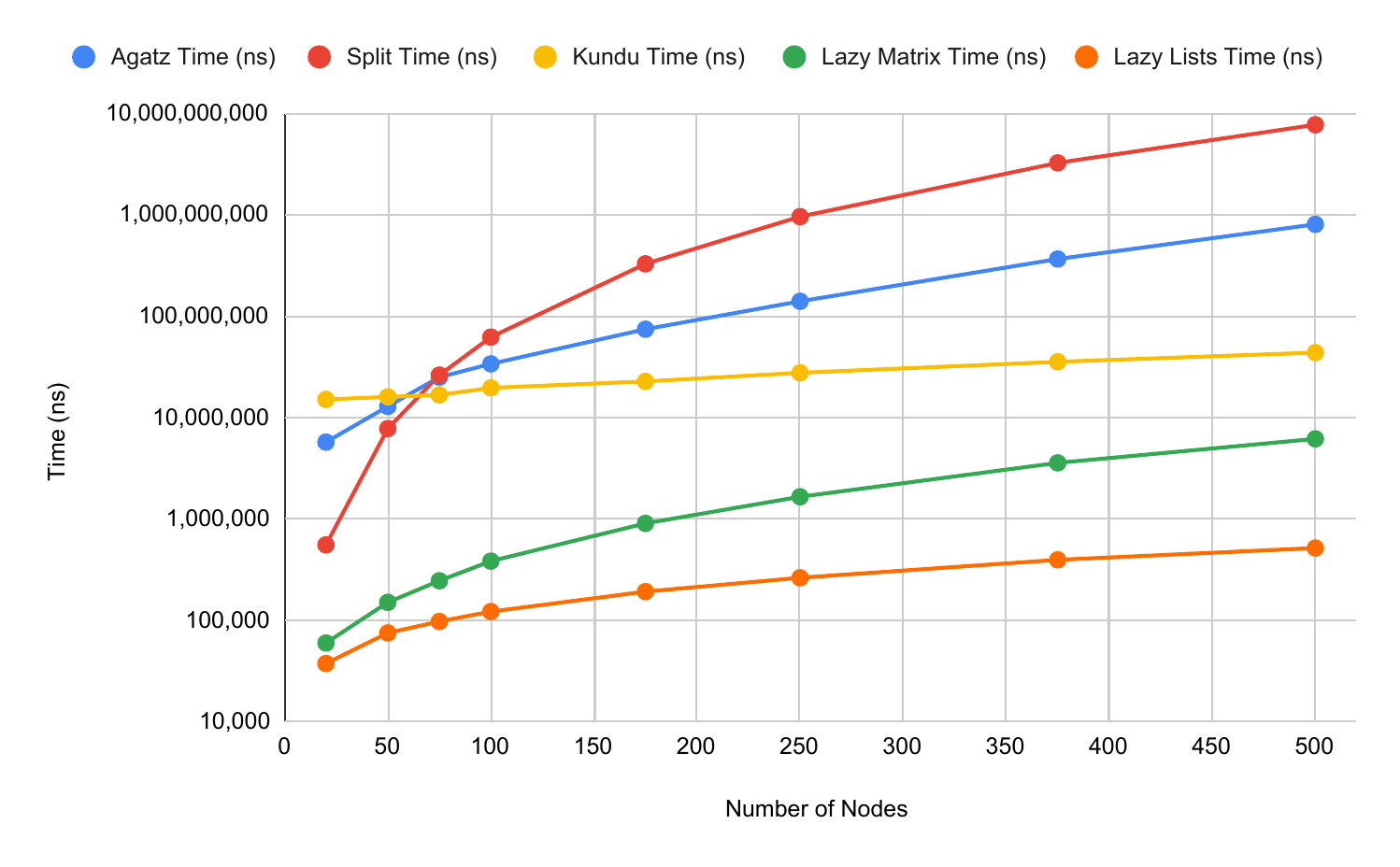}
    \caption{Visualizing average times of the algorithms for solving h-FSTSP from a cycle produced by LKH, as instances size increases.}
    \label{fig:times_lkh}
\end{figure}

Figure~\ref{fig:times_lkh} allows for a glimpse of the asymptotic behavior of each analyzed algorithm when operating on Hamiltonian cycles obtained by LKH.

The following analysis suggests that our algorithms are asymptotically faster in 
practice, which could lead to greater differences when dealing with larger instances. 

Table~\ref{tab:triples_lkh} shows the relation between the number of triples considered
by our lazy algorithms and the number of vertices of each instance.
The potential number of triples for an instance with~$n$ client nodes is ${n + 2 \choose 3} = \textrm{O}(n^3)$, since two nodes are used to represent the deposit.
This corroborates the high running times of the other algorithms in the literature.
However, Table~\ref{tab:triples_lkh} shows that the number of triples that need to be
considered by the lazy algorithms is proportional to the number of vertices, up to 
a small constant.
This helps to explain the performance of our lazy algorithms.
\begin{table}[!htb]
    \centering
    \caption{Relation between the number of triples considered by lazy algorithms and the number of nodes of each instance. The number of client nodes is $n$.}
    \label{tab:triples_lkh}
    \resizebox{0.45\textwidth}{!}{%
    \begin{tabular}{|c|c|c|c|c|c|c|c|}
    \hline
    & \multicolumn{2}{|c|}{Triples} & \multicolumn{2}{c|}{Triples / $(n+1)$} \\ \hline
    Number of Nodes & Average & Stdev & Average & Stdev \\ \hline\hline
    $\leq$ 10 & 15.01 & 12.31 & 1.95 & 1.37 \\ \hline
    20 & 54.81 & 47.70 & 2.74 & 2.39 \\ \hline
    50 & 149.64 & 135.11 & 2.99 & 2.70 \\ \hline
    75 & 234.41 & 224.04 & 3.13 & 2.99 \\ \hline
    100 & 307.68 & 288.70 & 3.08 & 2.89 \\ \hline
    175 & 519.29 & 471.62 & 2.97 & 2.69 \\ \hline
    250 & 747.74 & 686.61 & 2.99 & 2.75 \\ \hline
    375 & 1,099.82 & 994.18 & 2.93 & 2.65 \\ \hline
    500 & 1,462.36 & 1,311.33 & 2.92 & 2.62 \\ \hline
    \end{tabular}
    }
\end{table}

Furthermore, we noticed that the standard deviation of the ratio between 
the number of considered triples and the number of vertices is high in proportion to the average, 
which suggests that some characteristics of the instances can be determinants for the 
value of such ratio.
This led us to analyze the performance of the algorithms considering the~$\alpha$ factor of the instances, which indicates how many times the drone is faster than the truck.

Table~\ref{tab:alphas_lkh} shows the results of the algorithms when partitioning the instances by~$\alpha$ value and size.
The average reductions in solution costs are correlated with~$\alpha$, which is not surprising since a smaller~$\alpha$ means a slower drone. 
As expected, the running times of \textsc{Agatz} and \textsc{Split} are not affected by~$\alpha$. 
\textsc{Kundu} is faster for smaller~$\alpha$, which seems unusual.
\textsc{LazyMatrix} is faster for larger~$\alpha$, but its time is dominated by the cost to initialize the quadratic matrix.
Finally, \textsc{LazyLists} is much faster with larger~$\alpha$, which we conjecture to derive from the number of triples considered by the algorithm growing as~$\alpha$ decreases.
Moreover, the standard deviation of the ratio Triples / $(n+1)$ is very small when we partition the instances by~$\alpha$ value, showing the relevance of the drone speed over the impact of the Lazy Drone Property.

\begin{table}[!htb]
    \centering
    \caption{Results of the algorithms when partitioning the instances by $\alpha$ value and size.}
    \label{tab:alphas_lkh}
    \resizebox{\textwidth}{!}{%
    \begin{tabular}{|c|c|c|c|c|c|c|c|c|c|c|}
    \hline
    Number of &  & Number of & Cost & \textsc{Agatz} & \textsc{Split} & \textsc{Kundu} & \textsc{LazyMatrix} & \textsc{LazyLists} & \multicolumn{2}{c|}{Triples / $(n+1)$} \\ \cline{10-11}
    Nodes & \up{$\alpha$} & Instances & Reduction & Time (ns) & Time (ns) & Time (ns) & Time (ns) & Time (ns) & Average & Stdev  \\ \hline\hline
    $\leq$ 75 &  & 270 & -18.83\% & 7,914,191 & 3,823,131 & 14,553,554 & 89,934 & 58,065 & 4.70 & 1.66 \\ \cline{1-1}\cline{3-11}
    $\geq$ 100 and $\leq$ 250 & 1 & 120 & -18.23\% & 96,225,257 & 574,674,323 & 23,630,305 & 1,314,114 & 385,015 & 6.84 & 0.86 \\ \cline{1-1}\cline{3-11}
    $\geq$ 375 &  & 73 & -17.77\% & 547,606,588 & 5,118,728,082 & 37,099,093 & 5,092,980 & 836,574 & 6.62 & 0.50 \\ \hline\hline
    $\leq$ 75 &  & 270 & -29.34\% & 7,808,743 & 3,918,820 & 15,244,947 & 62,991 & 31,828 & 1.18 & 0.31 \\ \cline{1-1}\cline{3-11}
    $\geq$ 100 and $\leq$ 250 & 2 & 120 & -25.67\% & 96,403,239 & 572,785,045 & 23,969,931 & 1,056,607 & 124,479 & 1.15 & 0.07 \\ \cline{1-1}\cline{3-11}
    $\geq$ 375 &  & 74 & -25.25\% & 547,611,886 & 5,077,980,405 & 40,006,565 & 4,429,330 & 252,579 & 1.14 & 0.04 \\ \hline\hline
    $\leq$ 75 &  & 270 & -30.25\% & 7,789,698 & 3,829,865 & 15,823,628 & 62,134 & 30,414 & 0.98 & 0.18 \\ \cline{1-1}\cline{3-11}
    $\geq$ 100 and $\leq$ 250 & 3 & 120 & -26.25\% & 96,987,480 & 572,200,152 & 25,225,031 & 1,063,542 & 118,209 & 1.02 & 0.02 \\ \cline{1-1}\cline{3-11}
    $\geq$ 375 &  & 72 & -25.38\% & 545,750,210 & 5,071,019,166 & 39,004,848 & 4,395,283 & 243,671 & 1.02 & 0.02 \\ \hline
    \end{tabular}
    }
\end{table}

\begin{figure}[!htb]
    \centering
    \includegraphics[width=\textwidth]{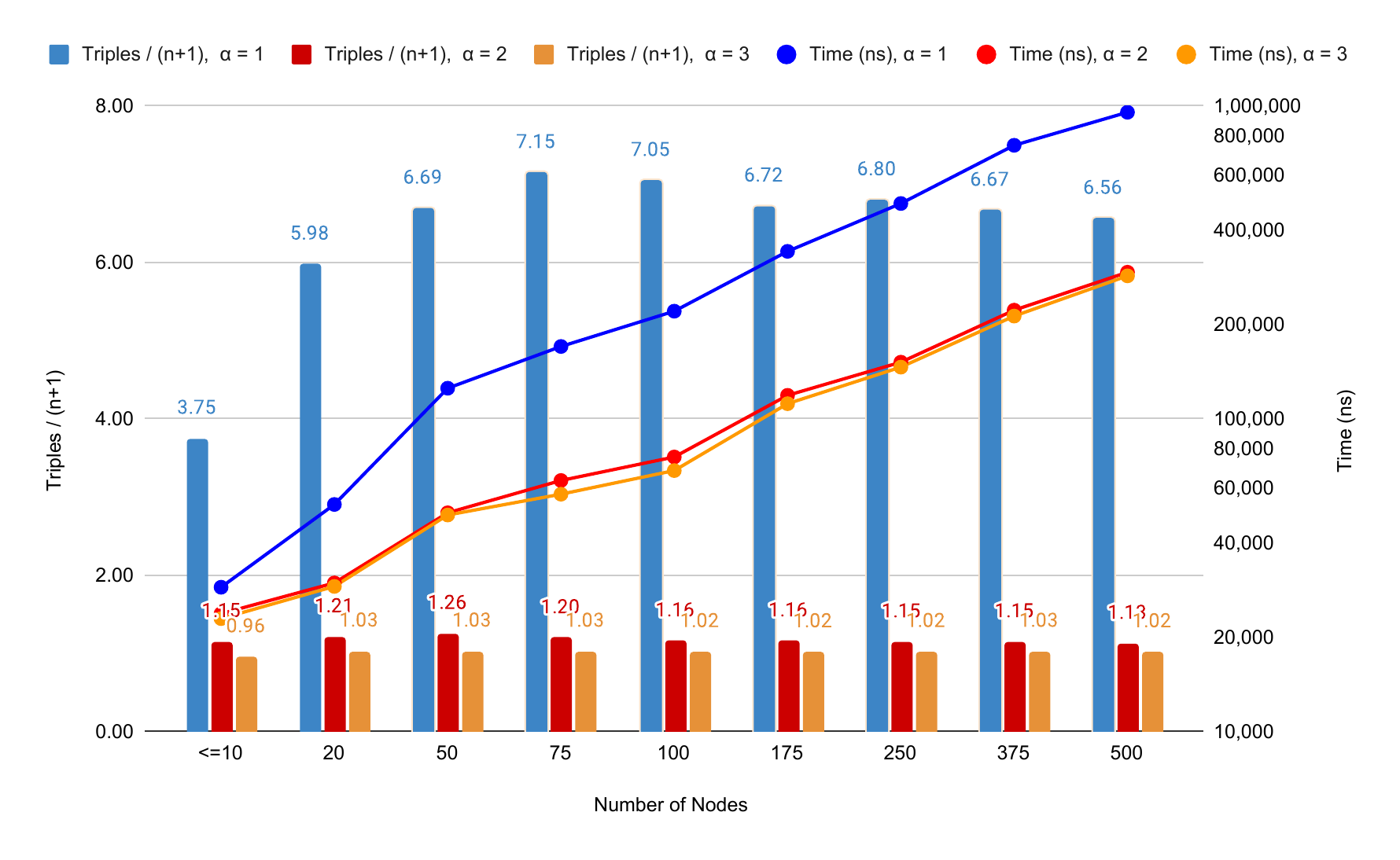}
    \caption{Correlation between \textsc{LazyLists} time and Triples / $(n+1)$ ratio according to $\alpha$ value and instances size.}
    \label{fig:alphas_lkh}
\end{figure}

Figure~\ref{fig:alphas_lkh} makes it easier to see the correlation between the running time of our best algorithm (\textsc{LazyLists}) and the number of triples per size as~$\alpha$ increases.

Similar conclusions can be taken when analyzing how each algorithm solved the h-FSTSP considering a Hamiltonian cycle produced by CHR and NNH. 
Their results are summarized in Tables~\ref{tab:CHR} and~\ref{tab:NNH}, respectively.
Despite deriving from different initial Hamiltonian cycles, the similar results achieved show that the effectiveness of the Lazy Drone Property and the behavior of the algorithms which use it are consistent.

\begin{table}[h!]
    \caption{Comparison among algorithms solving h-FSTSP from a cycle produced by CHR.}
    \label{tab:CHR}
    \centering
    \resizebox{\textwidth}{!}{%
    \begin{tabular}{|c|c|c|c|c|c|c|c|c|c|c|}
    \hline
    Number of &  & Number of & Cost & \textsc{Agatz} & \textsc{Split} & \textsc{Kundu} & \textsc{LazyMatrix} & \textsc{LazyLists} & \multicolumn{2}{c|}{Triples / $(n+1)$} \\ \cline{10-11}
    Nodes & \up{$\alpha$} & Instances & Reduction & Time (ns) & Time (ns) & Time (ns) & Time (ns) & Time (ns) & Average & Stdev  \\ \hline\hline

    $\leq$ 10 &  & 540 & -27.91\% & 4,434,891 & 59,853 & 10,566,625 & 30,771 & 25,339 & 1.96 & 1.35 \\ \cline{1-1}\cline{3-11}
    20 &  & 90 & -24.57\% & 5,640,849 & 544,307 & 10,382,334 & 55,632 & 37,129 & 2.73 & 2.31 \\ \cline{1-1}\cline{3-11}
    50 &  & 90 & -26.36\% & 12,157,992 & 7,751,115 & 11,382,202 & 141,681 & 72,102 & 3.08 & 2.75 \\ \cline{1-1}\cline{3-11}
    75 &  & 90 & -26.54\% & 24,480,287 & 25,916,103 & 12,052,825 & 249,757 & 102,839 & 3.21 & 2.94 \\ \cline{1-1}\cline{3-11}
    100 & Any & 90 & -25.17\% & 33,096,307 & 61,446,614 & 12,327,398 & 377,208 & 123,470 & 3.19 & 2.92 \\ \cline{1-1}\cline{3-11}
    175 & & 90 & -25.10\% & 72,701,267 & 327,378,122 & 14,420,379 & 904,688 & 200,125 & 3.08 & 2.77 \\ \cline{1-1}\cline{3-11}
    250 &  & 180 & -25.30\% & 135,393,153 & 952,791,806 & 17,185,549 & 1,663,806 & 280,515 & 3.14 & 2.87 \\ \cline{1-1}\cline{3-11}
    375 &  & 127 & -25.49\% & 355,187,884 & 3,220,993,543 & 21,962,579 & 3,556,141 & 409,445 & 3.05 & 2.72 \\ \cline{1-1}\cline{3-11}
    500 &  & 92 & -24.57\% & 783,215,296 & 7,658,545,435 & 25,562,482 & 6,078,664 & 543,563 & 3.03 & 2.67 \\ \hline\hline
    $\leq$ 75 &  & 270 & -19.95\% & 7,712,033 & 3,839,972 & 10,459,168 & 85,127 & 57,327 & 4.71 & 1.64 \\\cline{1-1}\cline{3-11}
    $\geq$ 100 and $\leq$ 250 & 1 & 120 & -19.92\% & 94,254,769 & 573,108,592 & 14,478,652 & 1,332,073 & 408,064 & 7.09 & 0.88 \\\cline{1-1}\cline{3-11}
    $\geq$ 375 &  & 73 & -19.77\% & 536,350,429 & 5,110,245,068 & 22,382,007 & 5,046,870 & 875,871 & 6.82 & 0.52 \\ \hline\hline
    $\leq$ 75 &  & 270 & -30.38\% & 7,595,601 & 3,849,467 & 10,908,856 & 62,232 & 33,103 & 1.20 & 0.31 \\\cline{1-1}\cline{3-11}
    $\geq$ 100 and $\leq$ 250 & 2 & 120 & -27.66\% & 93,482,885 & 573,762,714 & 15,591,431 & 1,049,728 & 132,504 & 1.24 & 0.09 \\\cline{1-1}\cline{3-11}
    $\geq$ 375 &  & 74 & -27.56\% & 535,256,582 & 5,082,150,541 & 24,206,090 & 4,444,289 & 275,282 & 1.23 & 0.06 \\ \hline\hline
    $\leq$ 75 &  & 270 & -31.31\% & 7,655,192 & 3,834,109 & 11,037,680 & 63,206 & 30,937 & 1.02 & 0.20 \\\cline{1-1}\cline{3-11}
    $\geq$ 100 and $\leq$ 250 & 3 & 120 & -28.07\% & 94,700,256 & 573,934,955 & 15,769,073 & 1,075,332 & 122,901 & 1.08 & 0.04 \\\cline{1-1}\cline{3-11}
    $\geq$ 375 &  & 72 & -27.99\% & 533,362,501 & 5,062,851,806 & 23,831,371 & 4,355,114 & 245,804 & 1.07 & 0.02 \\ \hline
    
    \end{tabular}
    }
\end{table}

\begin{table}[h!]
    \caption{Comparison among algorithms solving h-FSTSP from a cycle produced by NNH.}
    \label{tab:NNH}
    \centering
    \resizebox{\textwidth}{!}{%
    \begin{tabular}{|c|c|c|c|c|c|c|c|c|c|c|}
    \hline
    Number of &  & Number of & Cost & \textsc{Agatz} & \textsc{Split} & \textsc{Kundu} & \textsc{LazyMatrix} & \textsc{LazyLists} & \multicolumn{2}{c|}{Triples / $(n+1)$} \\ \cline{10-11}
    Nodes & \up{$\alpha$} & Instances & Reduction & Time (ns) & Time (ns) & Time (ns) & Time (ns) & Time (ns) & Average & Stdev  \\ \hline\hline

    $\leq$ 10 & & 540 & -29.04\% & 5,795,461 & 58,936 & 14,495,044 & 30,669 & 25,850 & 1.88 & 1.33 \\ \cline{1-1}\cline{3-11}
    20 &  & 90 & -25.84\% & 6,655,004 & 540,800 & 14,258,614 & 56,216 & 37,041 & 2.46 & 2.06 \\ \cline{1-1}\cline{3-11}
    50 &  & 90 & -26.66\% & 13,064,554 & 7,751,088 & 16,749,276 & 139,418 & 72,329 & 2.78 & 2.48 \\ \cline{1-1}\cline{3-11}
    75 &  & 90 & -23.98\% & 25,689,151 & 26,157,096 & 17,350,153 & 243,834 & 101,143 & 2.80 & 2.52 \\ \cline{1-1}\cline{3-11}
    100 & Any & 90 & -22.55\% & 35,824,467 & 61,499,041 & 19,479,848 & 372,884 & 119,442 & 2.84 & 2.58 \\ \cline{1-1}\cline{3-11}
    175 &  & 90 & -22.32\% & 76,164,949 & 327,691,111 & 24,202,757 & 889,037 & 201,745 & 2.85 & 2.57 \\ \cline{1-1}\cline{3-11}
    250 &  & 180 & -22.02\% & 140,756,869 & 960,615,561 & 28,001,483 & 1,639,260 & 267,003 & 2.83 & 2.55 \\ \cline{1-1}\cline{3-11}
    375 &  & 127 & -22.68\% & 361,980,779 & 3,236,189,370 & 36,282,383 & 3,504,620 & 382,554 & 2.82 & 2.53 \\ \cline{1-1}\cline{3-11}
    500 &  & 92 & -21.85\% & 798,069,745 & 7,689,411,413 & 47,120,386 & 6,002,346 & 506,673 & 2.83 & 2.52 \\ \hline\hline
    $\leq$ 75 &  & 270 & -21.19\% & 8,674,013 & 3,835,213 & 14,431,303 & 84,106 & 57,914 & 4.43 & 1.27 \\\cline{1-1}\cline{3-11}
    $\geq$ 100 and $\leq$ 250 & 1 & 120 & -17.46\% & 98,967,208 & 577,095,496 & 23,538,713 & 1,291,061 & 384,173 & 6.43 & 0.38 \\\cline{1-1}\cline{3-11}
    $\geq$ 375 &  & 73 & -17.69\% & 547,476,252 & 5,124,653,562 & 40,600,714 & 4,879,369 & 814,728 & 6.37 & 0.28 \\ \hline\hline
    $\leq$ 75 &  & 270 & -30.83\% & 9,097,857 & 3,839,963 & 15,336,741 & 62,909 & 32,342 & 1.08 & 0.19 \\\cline{1-1}\cline{3-11}
    $\geq$ 100 and $\leq$ 250 & 2 & 120 & -24.45\% & 98,429,785 & 577,692,648 & 25,498,731 & 1,053,129 & 134,857 & 1.09 & 0.04 \\\cline{1-1}\cline{3-11}
    $\geq$ 375 &  & 74 & -24.47\% & 543,087,994 & 5,106,614,189 & 41,457,884 & 4,436,115 & 248,776 & 1.10 & 0.02 \\ \hline\hline
    $\leq$ 75 &  & 270 & -31.57\% & 8,955,288 & 3,925,691 & 15,341,393 & 60,812 & 31,614 & 0.92 & 0.08 \\\cline{1-1}\cline{3-11}
    $\geq$ 100 and $\leq$ 250 & 3 & 120 & -24.76\% & 97,730,372 & 578,027,813 & 25,726,735 & 1,061,140 & 122,366 & 1.00 & 0.00 \\\cline{1-1}\cline{3-11}
    $\geq$ 375 &  & 72 & -24.84\% & 544,995,800 & 5,089,343,611 & 40,433,368 & 4,344,946 & 240,468 & 1.00 & 0.00 \\ \hline
    
    \end{tabular}
    }
\end{table}

%%%%%%%%%%%%%%%%%%%%%%%%%%%%%%%%%%%%%%%%%%%%%%%%%%%%%%%%%%%%%%%%%%%%%%%%%%%%%%%
\section{Final remarks}
\label{sec:remarks}

In this paper, we introduced the Lazy Drone property, which can be used to improve the practical 
performance of algorithms that solve the h-FSTSP and of order-first split-second heuristics for the FSTSP.
We also showed the Split Lazy Algorithm for h-FSTSP, which uses the property and yields running times
which are more than 84 times faster than the previously best-known algorithms when considering the literature benchmark.
Indeed, for the instances tested with more than 50 nodes, we observed that for each customer node, on average, only three pairs of launch and rendezvous nodes were considered.
We thus conclude that our algorithm presents a linear behavior in practice, which is quite divergent from
the cubic behavior of other algorithms.

It would be interesting now to prove the Lazy Drone property when considering other objective functions 
for the FSTSP and TSP-D problems, such as operational costs instead of completion time, as well as 
other restrictions for these problems, such as allowing the drone to be launched and retrieved 
at a same node and the flight endurance of the drone.
Also, it would be interesting to study in which conditions one can guarantee that the Lazy Drone property
is activated, which could allow us to present a theoretical guarantee of the efficiency of our algorithms.

\bibliographystyle{splncs04}
\bibliography{bibfile}

\begin{thebibliography}{10}
\providecommand{\url}[1]{\texttt{#1}}
\providecommand{\urlprefix}{URL }
\providecommand{\doi}[1]{https://doi.org/#1}

\bibitem{AgatzBS2018}
Agatz, N., Bouman, P., Schmidt, M.: {Optimization Approaches for the Traveling
  Salesman Problem with Drone}. {Transportation Science}  \textbf{52}(4),
  965--981 (2018). \doi{10.1287/trsc.2017.0791}

\bibitem{BellmoreN1968}
Bellmore, M., Nemhauser, G.L.: The traveling salesman problem: A survey.
  Operations Research  \textbf{16}(3),  538--558 (1968)

\bibitem{BoumanAS2018}
Bouman, P., Agatz, N., Schmidt, M.: {Instances for the TSP with Drone (and some
  solutions)} (2018). \doi{10.5281/zenodo.1204676}

\bibitem{Christofides1976}
Christofides, N.: Worst-case analysis of a new heuristic for the travelling
  salesman problem. Tech. rep., Carnegie-Mellon University. Pittsburgh,
  Pensylvania. Management Sciences Research Group (1976)

\bibitem{Christofides2022}
Christofides, N.: Worst-case analysis of a new heuristic for the travelling
  salesman problem. Operations Research Forum  \textbf{3}(1),  Paper No. 20, 4
  (2022). \doi{10.1007/s43069-021-00101-z}

\bibitem{HaDPH2018}
Ha, Q.M., Deville, Y., Pham, Q.D., H{\`a}, M.H.: {On the min-cost Traveling
  Salesman Problem with Drone}. {Transportation Research Part C: Emerging
  Technologies}  \textbf{86},  597--621 (2018). \doi{10.1016/j.trc.2017.11.015}

\bibitem{Helsgaun2000}
Helsgaun, K.: An effective implementation of the lin–kernighan traveling
  salesman heuristic. European Journal of Operational Research
  \textbf{126}(1),  106--130 (2000). \doi{10.1016/S0377-2217(99)00284-2}

\bibitem{Kundu2022}
Kundu, A.: Developing efficient order and split heuristics for coordinated
  covering tour problems with drones. Ph.D. thesis, Texas Tech University
  (2022), \url{https://hdl.handle.net/2346/89264}

\bibitem{KunduEM2022}
Kundu, A., Escobar, R.G., Matis, T.I.: An efficient routing heuristic for a
  drone-assisted delivery problem. {IMA Journal of Management Mathematics}
  \textbf{33}(4),  583--601 (2022). \doi{10.1093/imaman/dpab039}

\bibitem{LinK1973}
Lin, S., Kernighan, B.W.: An effective heuristic algorithm for the
  traveling-salesman problem. Operations Research  \textbf{21}(2),  498--516
  (1973). \doi{10.1287/opre.21.2.498}

\bibitem{MacrinaDGL2020}
Macrina, G., {Di Puglia Pugliese}, L., Guerriero, F., Laporte, G.: {Drone-aided
  routing: A literature review}. {Transportation Research Part C: Emerging
  Technologies}  \textbf{120},  102762 (2020). \doi{10.1016/j.trc.2020.102762}

\bibitem{MurrayC2015}
Murray, C.C., Chu, A.G.: {The flying sidekick traveling salesman problem:
  Optimization of drone-assisted parcel delivery}. {Transportation Research
  Part C: Emerging Technologies}  \textbf{54},  86--109 (2015).
  \doi{10.1016/j.trc.2015.03.005}

\end{thebibliography}
    
\end{document}